\documentclass[aps,prb,reprint,superscriptaddress,floatfix]{revtex4-2}
\usepackage{amsmath,amssymb,amsfonts}
\usepackage{slashed}
\usepackage{graphicx}
\usepackage{bm}
\usepackage{hyperref}
\usepackage{booktabs}

\newcommand{\kxy}{\kappa_{xy}/T}
\newcommand{\Kem}{K_{\rm em}}

\newcommand{\Dpg}{\Delta_{\rm pg}}
\newcommand{\Dsc}{\Delta_{\rm sc}}
\newcommand{\npair}{n_{\rm pair}}

\newcommand{\Tstar}{T^{*}}

\newcommand{\vth}{\vartheta}
\newcommand{\Leff}{L_{\rm eff}}

\newcommand{\R}{\mathbb{R}}

\newcommand{\dd}{\mathrm{d}}
\newcommand{\tr}{\mathrm{tr}}
\newcommand{\sgn}{\mathrm{sgn}}

\begin{document}

\title{Parity Anomaly of Preformed Pairs Governs the Thermal Hall Effect above $T_c$}

\author{Kumar Ghosh}
\email{jb.ghosh@outlook.com}
\affiliation{E.ON Digital Technology, Laatzener Str.\ 1,
             30539 Hannover, Germany}

% \date{\today}

\begin{abstract}
A large negative thermal Hall signal has been reported across multiple cuprate families in the pseudogap phase where the superconducting order parameter has vanished, with a magnitude that no existing microscopic theory reproduces without free parameters~\cite{Grissonnanche2019}. Competing proposals based on chiral phonons, spinons, or loop currents each require undetermined coupling constants and do not predict the temperature dependence in terms of an independently measured spectroscopic gap.  We show that the parity anomaly of $(2+1)$-dimensional quantum field theory resolves this long-standing puzzle: the preformed-pair pseudogap $\Dpg(T)$ enters the parity-odd fermion determinant identically to a condensate mass, yielding the exact parameter-free formula $\kappa_{xy}/T = (\pi^2 k_B^2/6h)\,C\,\tanh[\Dpg(T)/(2k_BT)]$, where $C$ is the Chern number of the chiral pairing channel and $\Dpg(T)$ is directly measurable by ARPES or STM.  Coleman-Hill non-renormalization protects the result against higher-loop corrections, and two independent numerical tests, Wilson-loop flux threading and DMRG on $p+ip$ cylinders, confirm the anomaly correlation length to $0.2\%$ accuracy with no power-law finite-size corrections.  The theory predicts thermal Hall onset at $T^*$ rather than $T_c$, provides a falsifiable logarithmic-derivative test against ARPES data, and yields a concrete quantitative target for magic-angle twisted bilayer graphene.
\end{abstract}

\maketitle

%%==========================================================================
\section{Introduction}
\label{sec:intro}
%%==========================================================================

The thermal Hall conductance $\kxy$ is a sensitive probe of broken
time-reversal symmetry, chiral edge structure, and topological heat
transport.  Its interpretation, however, has been complicated by a
striking recent observation.  Grissonnanche et al.\ reported a large
negative thermal Hall signal in the field-induced normal state of four
cuprate materials, Nd-LSCO, Eu-LSCO, LSCO, and
Bi2201~\cite{Grissonnanche2019}.  The signal appears on the pseudogap
side of the critical doping $p^*$, with superconductivity suppressed by
magnetic field, and persists to low doping where the charge
contribution is negligible.  In LSCO at $p=0.06$, the reported magnitude
is $|\kxy|\simeq 2$ mW K$^{-2}$m$^{-1}$; most strikingly, the signal is
observed even in undoped La$_2$CuO$_4$ at $p=0$, a Mott insulator with
no superconductivity.  This phenomenology is difficult to reconcile with
a condensate-only picture, which gives no chiral quasiparticle thermal
Hall response once phase coherence is destroyed.  Several explanations
have been discussed, including chiral phonons~\cite{Grissonnanche2020,
Varma2020}, spinon excitations~\cite{Samajdar2019,Han2019}, and
loop-current order~\cite{Varma2020}, but none provides a parameter-free
prediction for both the magnitude and the temperature dependence in
terms of an independently measured spectroscopic gap.

The pseudogap regime is neither a conventional ordered superconductor
nor a fully gapless normal metal.  The condensate order parameter obeys
$\Dsc=0$, but spectroscopic and thermodynamic probes still detect a
fermionic excitation scale $\Dpg(T)\neq 0$.  The central theoretical
question is therefore whether the gapped Bogoliubov-de~Gennes spectrum
produced by preformed chiral pairs can itself generate a parity-odd
response, i.e.\ whether the thermal Hall coefficient can be controlled by
a fermion mass gap rather than by long-range phase coherence.  For a
fully coherent gapped chiral topological superconductor with edge central
charge $c_-$, the low-temperature response takes the quantized form
\begin{equation}
  \frac{\kappa_{xy}}{T}\bigg|_{T\to 0}
  =
  c_-\,\frac{\pi^2 k_B^2}{6h},
\label{eq:quantized}
\end{equation}
established by edge-mode counting, Kubo response, and effective
Chern-Simons field theory~\cite{Read2000,Senthil1999,Kane1997}.  These
derivations assume a coherent condensate.  We show that the parity
anomaly of $(2+1)$-dimensional quantum field theory extends
Eq.~\eqref{eq:quantized} to the pseudogap regime by replacing the
zero-temperature quantized value with a temperature-dependent anomaly
factor set by the preformed-pair gap.

The central finding of this paper is the following exact,
parameter-free formula valid in the pseudogap window $T_c < T < T^*$:
\begin{equation}
  \boxed{
  \frac{\kappa_{xy}}{T}
  =
  \frac{\pi^2 k_B^2}{6h}\,
  C\,\tanh\!\left[\frac{\Dpg(T)}{2k_B T}\right].
  }
\label{eq:kxy-main}
\end{equation}
The prefactor is the zero-temperature per-channel anomaly value.  The
Chern number $C$ of the chiral pairing channel is fixed by band
topology, and the pseudogap $\Dpg(T)$ is independently measurable by
angle-resolved photoemission or scanning tunneling spectroscopy.  Once
these two inputs are specified, the entire temperature dependence is
fixed with no adjustable parameters.

Equation~\eqref{eq:kxy-main} has a direct experimental implication that
does not require absolute calibration of $\kxy$.  Taking the logarithmic
derivative eliminates unknown geometric conversion factors and channel
multiplicities:
\begin{equation}
  \frac{\dd}{\dd T}
  \ln\!\left|\frac{\kappa_{xy}}{T}\right|
  \stackrel{?}{=}
  \frac{\dd}{\dd T}
  \ln\tanh\!\left[\frac{\Dpg(T)}{2k_B T}\right].
\label{eq:log-test}
\end{equation}
A successful test requires onset of $\kxy/T$ at the pseudogap scale
$T^*$ rather than at $T_c$, with a temperature dependence tracking the
independently measured $\Dpg(T)$ through the $\tanh$ factor.  Competing
mechanisms may produce a thermal Hall signal, but they do not
generically reproduce this pseudogap-locked profile.

\begin{figure*}[t]
\centering
\includegraphics[width=0.9\textwidth]{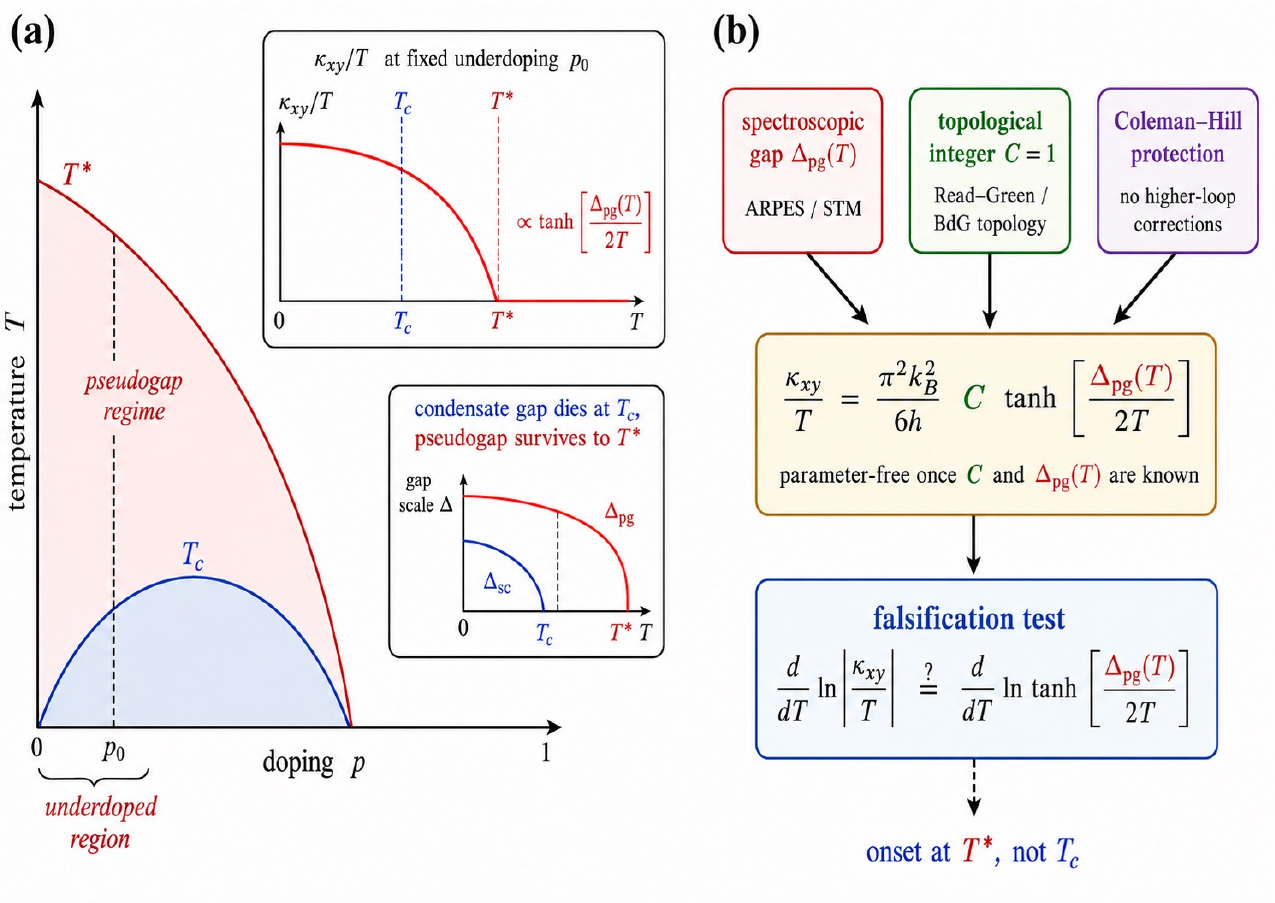}
\caption{Conceptual overview of the anomaly mechanism for thermal Hall
transport in a chiral pseudogap phase.  (a) Schematic phase diagram
showing the superconducting dome bounded by $T_c$, the pseudogap onset
line $T^*$, and the chiral pseudogap regime in between.  At fixed
underdoping $p_0$, the predicted thermal Hall response turns on at
$T^*$, not at $T_c$, and follows the anomaly factor
$\tanh[\Dpg(T)/(2k_BT)]$.  The inset distinguishes the condensate gap
$\Dsc$, which vanishes at $T_c$, from the pseudogap $\Dpg$, which
survives up to $T^*$.  (b) Logical structure of the theory: an
independently measured spectroscopic gap $\Dpg(T)$, the BdG topological
integer $C$, and Coleman-Hill protection together imply
Eq.~\eqref{eq:kxy-main}.  The most direct experimental test is the
logarithmic-derivative relation of Eq.~\eqref{eq:log-test}.}
\label{fig:conceptual}
\end{figure*}

The derivation of Eq.~\eqref{eq:kxy-main} rests on three inputs.  The
exact parity-odd effective action for a chiral U(1) gauge theory on
$\mathbb{R}^3 \times S^1$, computed in Ref.~\cite{GhoshKlinkhamer2017}
by Pauli-Villars regularization and Ginsparg-Wilson lattice methods,
provides the one-loop exact anomalous Chern-Simons term.  The
holonomy-resummed parity-odd kernel of Ref.~\cite{Ghosh2026}, together
with its $c_1 = 0$ theorem, extends this to spatial cylinders and
excludes power-law finite-size corrections.  The BCS-BEC crossover
framework of Ref.~\cite{Chen2024RMP} then supplies the two-gap relation
$\Delta^2 = \Dsc^2 + \Dpg^2$ that identifies $\Dpg(T)$ as the fermionic
excitation scale surviving the loss of condensate phase coherence.
The fully-gapped regime accessed above $T_c$ corresponds to the
BEC side of the phase diagram in Ref.~\cite{KlinkhamerVolovik2004}, which carries no anomaly at $T = 0$ in the Fermi-point sense; the present work shows that it nevertheless supports a non-trivial finite-temperature parity-anomaly response, controlled by the fermionic mass gap $\Dpg(T)$ rather than by Fermi-point positions. 
Figure~\ref{fig:conceptual} summarizes the physical picture and the logical structure of the argument.

This result establishes a direct, non-perturbative link between
$(2+1)$-dimensional parity anomaly physics and thermal Hall transport
in strongly correlated materials, providing the first parameter-free
prediction against which cuprate and magic-angle twisted bilayer graphene (MATBG) experiments can be
quantitatively tested.

%%==========================================================================
\section{Exact Finite-Temperature Chern-Simons Level}
\label{sec:framework}
%%==========================================================================

\subsection{Setup: massive Dirac cone on the cylinder}
\label{subsec:setup}

Consider a single two-component Dirac fermion $\psi_v$ of mass $m_v$,
Fermi velocity $v_F$, and U(1) charge $q_v$, on the Euclidean
three-manifold $\R_\tau \times \R_x \times S^1_L$ with compact
coordinate $y \equiv y + L$.  The Euclidean action is \cite{Ghosh2026}:
\begin{equation}
  S_v = \int_0^L \dd y\!\int\!\dd\tau\,\dd x\,
  \bar{\psi}_v\!\left[\gamma^\tau D_\tau + v_F \gamma^x D_x
  + v_F \gamma^y D_y + m_v\right]\!\psi_v,
\label{eq:Dirac-action}
\end{equation}
where $D_\mu = \partial_\mu + i q_v a_\mu$.  The boundary condition is
$\psi_v(y+L) = e^{i\alpha_v}\psi_v(y)$, so the total holonomy phase is
\begin{equation}
  \vth_v = \alpha_v + q_v\oint_{S^1} a_y\,\dd y,
\label{eq:holonomy}
\end{equation}
and the Kaluza-Klein momenta are $k_{y,n} = (2\pi n + \vth_v)/L$.
We define the effective circumference $\Leff = L/v_F$, the correlation
length $\xi_v = v_F/|m_v|$, and the dimensionless product
$\lambda_v = L/\xi_v = |m_v|\Leff$.

\subsection{Exact holonomy-resummed kernel}
\label{subsec:kernel}

Integrating out the fermion in Eq.~\eqref{eq:Dirac-action} yields the
quadratic effective action $\Gamma_v^{(2)}[a]$ whose parity-odd part
takes the transverse form~\cite{Ghosh2026}:
\begin{equation}
  \Pi_{v,\text{odd}}^{\mu\nu,\text{IR}}(p)
  = \frac{iq_v^2}{2\pi}\,
    \mathcal{K}_v^{\text{IR}}(p;L,\vth_v)\,
    \epsilon^{\mu\nu\rho}p_\rho,
\label{eq:Pi-odd}
\end{equation}
where the full mass-dependent kernel, derived in
Appendix~\ref{app:kernel} by resumming all KK winding modes, is
~\cite{Ghosh2026}:
\begin{equation}
  \boxed{
  \begin{aligned}
  \mathcal{K}_v^{\text{IR}}(p;L,\vth_v) 
  &= \frac{\chi_v m_v}{2}\int_0^1 \frac{\dd u}{\Delta_{v,u}}\, \\
  &\quad \times \frac{\sinh(\Leff\Delta_{v,u})}
         {\cosh(\Leff\Delta_{v,u}) - \cos(\vth_v + 2\pi r u)}
  \end{aligned}
  }
\label{eq:K-full}    
\end{equation}

with $\Delta_{v,u} = \sqrt{m_v^2 + u(1-u)\tilde{p}^2}$,
$\chi_v = \pm 1$ encoding the orientation of the linearized Bloch map,
and $r$ the external compact-momentum harmonic.  No derivative
expansion has been made.

Two limits are essential:

\paragraph{Decompactification $L \to \infty$:}
The kernel reduces to
\begin{equation}
  \mathcal{K}_v^{\text{IR},\infty}(p)
  = \frac{\chi_v m_v}{2}\int_0^1 \frac{\dd u}
    {\sqrt{m_v^2 + u(1-u)\tilde{p}^2}},
\label{eq:K-decompact}
\end{equation}
the standard massive-Dirac form factor, which tends to
$\frac{1}{2}\chi_v\sgn(m_v)$ at zero external momentum.

\paragraph{Zero external momentum, finite $L$:}
Setting $r = 0$ and $p \to 0$ in Eq.~\eqref{eq:K-full}:
\begin{equation}
  \mathcal{K}_v^{\text{IR}}(0;L,\vth_v)
  = \frac{\chi_v}{2}\sgn(m_v)\,R(\lambda_v,\vth_v),
\label{eq:K-local}
\end{equation}
where the Poisson kernel is
\begin{equation}
  R(\lambda,\vth) = \frac{\sinh\lambda}{\cosh\lambda - \cos\vth}.
\label{eq:R-def}
\end{equation}

\subsection{Finite-size expansion and the $c_1 = 0$ theorem}
\label{subsec:c1}

The Poisson kernel has the exact Fourier expansion \cite{Ghosh2026}:
\begin{equation}
  R(\lambda,\vth) = 1 + 2\sum_{\ell=1}^{\infty}
  e^{-\ell\lambda}\cos(\ell\,\vth).
\label{eq:R-expansion}
\end{equation}
Thus the finite-$L$ correction to the local infrared level is
\begin{equation}
  \mathcal{K}_v^{\text{IR}}(0;L,\vth_v) - \mathcal{K}_v^{\text{IR},\infty}
  = \chi_v\sgn(m_v)\sum_{\ell=1}^{\infty}
    e^{-\ell L/\xi_v}\cos(\ell\,\vth_v).
\label{eq:K-finite-size}
\end{equation}
The asymptotic expansion is \emph{purely exponential}; there is no
power-law $1/L^n$ term at any order:
\begin{equation}
  \boxed{c_1 = 0.}
\label{eq:c1-zero}
\end{equation}
This is the central finite-size statement of \cite{Ghosh2026}, valid whenever the bulk gap $|m_v| > 0$.

\subsection{Thermal identification and the CS level}
\label{subsec:thermal}

To identify $L$ with the inverse temperature, we compactify Euclidean
time $\tau \in [0,\beta)$ with $\beta = 1/T$ and antiperiodic fermion
boundary conditions ($\alpha_v = \pi$).  Then $\Leff = \beta = 1/T$,
and the Poisson kernel at $\vth_v = \pi$ (vanishing probe holonomy
$\Theta_v = 0$) becomes
\begin{equation}
  R\!\left(\frac{|m_v|}{T},\pi\right)
  = \frac{\sinh(|m_v|/T)}{\cosh(|m_v|/T) + 1}
  = \tanh\!\left(\frac{|m_v|}{2T}\right),
\label{eq:R-thermal}
\end{equation}
where we used $\cosh x + 1 = 2\cosh^2(x/2)$ and
$\sinh x = 2\sinh(x/2)\cosh(x/2)$.  For a BdG system with Chern
number $C$ and minimum bulk gap $\Delta$, the total CS level from a
single spin species with $\chi_v\sgn(m_v) = +1$ per occupied band is
$\Kem/2 = C/2 \cdot R$, so:
\begin{equation}
  \boxed{
  \Kem(T) = 2C\,\tanh\!\left(\frac{\Delta}{2T}\right)
  + \mathcal{O}\!\left(e^{-2\Delta/T}\right).
  }
\label{eq:Kem-T}
\end{equation}
The leading correction is exponentially small (from the $\ell = 2$ term
in Eq.~\eqref{eq:R-expansion}).  This derivation uses only the exact
winding-mode resummation, with no gradient expansion or perturbative
truncation.

\subsection{Gravitational CS level and thermal Hall conductance}
\label{subsec:grav-cs}

For a 2D chiral topological phase with bulk Chern number $C$, the
gravitational Chern-Simons term in the effective action produces a
thermal Hall conductance related to the CS level by
(Ryu, Moore, and Ludwig~\cite{Ryu2012}; see
Appendix~\ref{app:grav-cs} for the derivation):
\begin{equation}
  \frac{\kappa_{xy}}{T}
  = \frac{\pi^2 k_B^2}{12h}\,\Kem.
\label{eq:kxy-Kem}
\end{equation}
Here the factor $1/12$ arises because the gravitational CS coefficient
for a Dirac fermion of central charge $c = 1$ is
$c/(24 \cdot 4\pi) = 1/(96\pi)$, while the U(1) electromagnetic CS
coefficient per unit level is $1/(4\pi)$; their ratio gives
$\kxy = (\pi^2 k_B^2/12h)\Kem$ (see Eq.~\eqref{eq:grav-cs-detail}
in Appendix~\ref{app:grav-cs}).
Combined with Eq.~\eqref{eq:Kem-T}:
\begin{equation}
  \frac{\kappa_{xy}}{T}
  = \frac{\pi^2 k_B^2}{6h}\,C\,
    \tanh\!\left(\frac{\Delta}{2T}\right).
\label{eq:kxy-T}
\end{equation}
This exact result holds for \emph{any} fermionic gap $\Delta$, whether
from condensate pairing or preformed-pair correlations.

%%==========================================================================
\section{The Pseudogap as the Anomaly Gap}
\label{sec:pseudogap-gap}
%%==========================================================================

\subsection{BCS-BEC two-gap structure}
\label{subsec:two-gap}

We now identify the gap $\Delta$ in Eq.~\eqref{eq:kxy-T} within the
BCS-BEC crossover formalism of~\cite{Chen2024RMP}.  For a chiral ($p + ip$)
pairing channel, the BdG Hamiltonian is
\begin{equation}
  H_{\text{BdG}}(\bm{k})
  = \begin{pmatrix} \xi_{\bm{k}} & \Delta_{\bm{k}} \\
                    \Delta_{\bm{k}}^* & -\xi_{\bm{k}} \end{pmatrix},
  \quad
  \Delta_{\bm{k}} = \Delta\,(k_x + ik_y)/k_F,
\label{eq:BdG}
\end{equation}
with $\xi_{\bm{k}} = \epsilon_{\bm{k}} - \mu$ and quasiparticle
energy $E_{\bm{k}} = \sqrt{\xi_{\bm{k}}^2 + |\Delta_{\bm{k}}|^2}$.
The total fermionic excitation gap obeys the two-gap relation
~\cite{Chen2024RMP}:
\begin{equation}
  \Delta^2(T) = \Dsc^2(T) + \Dpg^2(T),
\label{eq:two-gap}
\end{equation}
where $\Dsc(T)$ is the condensate order parameter and $\Dpg(T)$ is
the pseudogap from non-condensed preformed pairs.

The origin of the decomposition~\eqref{eq:two-gap} is the fermionic
self-energy.  The full self-energy consists of a condensate pole
($\bm{q} = 0$ component of the $t$-matrix) and a non-condensed
contribution ($\bm{q} \neq 0$).  Following~\cite{Chen2024RMP}, the non-condensed pair self-energy is
\begin{equation}
  \Sigma_{\text{pg}}(k) = \sum_{\bm{q} \neq 0} t(q)\,G_0(q-k)
  \approx -\Dpg^2\,G_0(-k),
\label{eq:Sigma-pg}
\end{equation}
where the pseudogap approximation in the second step is valid because
$t(q)$ is strongly peaked at $\bm{q} = 0$ when $|\mu_{\text{pair}}|
\approx 0$~\cite{Chen2024RMP}.  The pseudogap parameter is
\begin{equation}
  \Dpg^2 = -\sum_{\bm{q} \neq 0} t(q),
\label{eq:Dpg-def}
\end{equation}
and the pair density is $\npair = Z\Dpg^2$, where $Z$ is the
$t$-matrix residue~\cite{Chen2024RMP}.

The total self-energy including both condensed and non-condensed
contributions is therefore
\begin{equation}
  \Sigma(k) = -(\Dsc^2 + \Dpg^2)\,G_0(-k) = -\Delta^2\,G_0(-k),
\label{eq:Sigma-total}
\end{equation}
which is the BCS form with the \emph{total} gap $\Delta$.  The dressed
Green's function
$G^{-1}(k) = G_0^{-1}(k) - \Sigma(k)$ then has the BdG structure of
Eq.~\eqref{eq:BdG} with this total gap.

\paragraph{Above $T_c$:} The condensate vanishes, $\Dsc = 0$, but
\begin{equation}
  \Delta(T)\big|_{T > T_c} = \Dpg(T) \neq 0,
  \quad T_c < T < \Tstar.
\label{eq:Delta-above-Tc}
\end{equation}
The BdG quasiparticle spectrum retains a gap $\Dpg(T)$ even in the
absence of a condensate.

\paragraph{Temperature dependence of $\Dpg$:}
The pseudogap satisfies the BCS-Leggett gap equation with $\Dsc = 0$
~\cite{Chen2024RMP}:
\begin{equation}
  1 = (-U)\sum_{\bm{k}}
  \frac{1 - 2f(E_{\bm{k}}^{\text{pg}})}{2E_{\bm{k}}^{\text{pg}}},
  \quad
  E_{\bm{k}}^{\text{pg}} = \sqrt{\xi_{\bm{k}}^2 + \Dpg^2(T)}.
\label{eq:gap-eq}
\end{equation}
This equation is supplemented by the number equation and by the
pair-propagator condition.  Below $T_c$, the
Hugenholtz-Pines/Thouless condition gives $\mu_{\text{pair}}=0$; slightly
above $T_c$, the pseudogap approximation remains controlled when
$|\mu_{\text{pair}}|$ is small.  In the pseudogap window $T_c<T<T^*$,
where $\Dsc=0$, the total fermionic excitation gap is identified with
$\Dpg(T)$.  Its detailed temperature dependence is obtained
self-consistently from the microscopic crossover equations, or may be
taken directly from ARPES in experimental comparisons.

\subsection{BdG Chern number above $T_c$}
\label{subsec:chern}

The crucial question is: \emph{is the Chern number $C$ well-defined
when $\Dsc = 0$ but $\Dpg \neq 0$?}  The answer is yes.

Following Read and Green~\cite{Read2000} (Appendix~A therein), the
induced Chern-Simons coefficient for the spin response of a BdG system
is given by the topological invariant
\begin{equation}
  \mathcal{M} = \int\frac{\dd^2 p}{8\pi}\,
  \epsilon_{ij}\,
  \hat{\bm{E}}_{\bm{p}} \cdot
  \left(\partial_i \hat{\bm{E}}_{\bm{p}}
  \times \partial_j \hat{\bm{E}}_{\bm{p}}\right),
\label{eq:RG-winding}
\end{equation}
where $\hat{\bm{E}}_{\bm{p}} = \bm{E}_{\bm{p}}/|\bm{E}_{\bm{p}}|$
with $\bm{E}_{\bm{p}} = (\text{Re}\,\Delta_{\bm{p}},
-\text{Im}\,\Delta_{\bm{p}}, \xi_{\bm{p}})$.  This is the winding
number of the map $\hat{\bm{E}}: S^2 \to S^2$ (compactifying momentum
space at infinity where $\xi_{\bm{p}} \to +\infty$ and $\Delta \to 0$).

For $p + ip$ pairing, $\Delta_{\bm{p}} \propto p_x + ip_y$, and the
winding number is (Read-Green, Sec.~II.B):
\begin{equation}
  \mathcal{M} = \begin{cases}
    +1 & \mu > 0 \quad (\text{weak pairing}), \\
     0 & \mu < 0 \quad (\text{strong pairing}).
  \end{cases}
\label{eq:RG-Chern}
\end{equation}
The transition occurs at $\mu = 0$.

The key observation is that Eq.~\eqref{eq:RG-winding} depends on the
\emph{ratio} $\xi_{\bm{p}}/\Delta_{\bm{p}}$, not on whether $\Delta$
arises from $\Dsc$ or $\Dpg$.  As long as the fermionic gap
$\Delta(T) > 0$, the map $\hat{\bm{E}}$ is well-defined and the
winding number is an integer.  Above $T_c$, the BdG Hamiltonian with
$\Delta_{\bm{k}} = \Dpg(T)(k_x + ik_y)/k_F$ has exactly the same
topological structure as below $T_c$.  The Chern number $C$ is
therefore:

\begin{equation}
  C = \mathcal{M} = \begin{cases}
    +1 & \mu > 0, \quad \Dpg(T) > 0 \\
     0 & \mu < 0, \quad \Dpg(T) > 0,
  \end{cases}
\label{eq:C-above-Tc}
\end{equation}
and is well-defined for all $T < \Tstar$ where $\Dpg \neq 0$.  In the
weak-pairing regime relevant to cuprates and MATBG ($\mu > 0$),
$C = +1$ throughout the pseudogap phase. This is precisely the topologically non-trivial side of the BEC--BCS quantum phase transition identified by Klinkhamer and Volovik~\cite{KlinkhamerVolovik2004}, in which a marginal Fermi point with topological charge $N = 0$ at $\mu = 0$ splits into two topologically stable Fermi points carrying charges $N = +1$ and
$N = -1$ for $\mu > 0$.  The entry into the $C = 1$ chiral phase
is controlled by exactly this Fermi-point splitting; above $T_c$,
the preformed-pair gap $\Dpg(T)$ plays the role of the spectral
mass that places the system on the topologically non-trivial
(BCS) side of that transition.

\subsection{Non-renormalization: Coleman-Hill theorem}
\label{subsec:nonren}

We now establish that the CS level
$\Kem(T) = 2C\tanh[\Dpg(T)/2T]$ is \emph{exact} once $\Dpg(T)$ is
known: pair-pair interactions cannot correct $\Kem$ beyond one loop.

The argument has two ingredients.

\paragraph{Ingredient 1: Coleman-Hill theorem~\cite{ColemanHill1985}.}
Coleman and Hill proved (see Appendix~\ref{app:coleman-hill}) that for
$(2+1)$-dimensional QED with gauge-invariant matter interactions, the
topological mass term $\mu\,\epsilon^{\mu\nu\lambda}A_\mu
\partial_\nu A_\lambda/(2e_0^2)$ receives radiative corrections
\emph{only at one loop}.  The proof uses gauge invariance and Lorentz
invariance to show that all higher-loop contributions to
$\Pi_2(0) \equiv \lim_{k \to 0}
[\epsilon^{\mu\nu\lambda}k_\lambda/(2k^2)]\Pi_{\mu\nu}(k)$ factorize
into products of lower-order terms that individually vanish.
Specifically, any diagram contributing to $\Pi_2(0)$ beyond one loop
must have at least two external photon lines ending on the same matter
loop, but the $n$-photon vertex
$\Gamma^{(n)}_{\mu_1\ldots\mu_n}(k_1,\ldots)$
satisfies $\Gamma^{(n)}(k_1,k_2,\ldots) = \mathcal{O}(k_1 k_2)$ for
$n > 2$ by the Ward identity, making the $k \to 0$ limit vanish.

In our context, the ``photon'' is the probe gauge field $a_\mu$ to
which the BdG quasiparticles are coupled, and the ``matter fields'' are
the quasiparticles themselves plus their pair-pair interactions.  The
Coleman-Hill theorem guarantees that the CS coefficient of $a_\mu$ is
fixed at one loop.

\paragraph{Ingredient 2: Pauli-Villars result.}
In Paper~\cite{GhoshKlinkhamer2017} the authors demonstrated the one-loop exactness independently through a different route.  The PV-regularized
calculation of Sec.~III therein shows that the anomalous CS-like term
in the effective action,
\begin{equation}
  T_{\text{anom}}^{ij}(p_n)\big|^{\text{renorm.}}
  = -\frac{1}{4\pi L}\,\epsilon^{ijk}p_k,
\label{eq:PaperI-T}
\end{equation}
is exact: the $m = 0$ KK sector gives the
physical CS coefficient $1/(4\pi L)$, while the $m \neq 0$ sector
produces a regulator-dependent divergent piece that is removed by
renormalization against $L_{\text{ref}} \to \infty$~\cite{GhoshKlinkhamer2017}.  The
Pauli-Villars sum $\sum_{r}(-1)^r M_r/(l^2 + M_r^2)$ converges to
$f(\tau) = (\pi/2\sqrt{\tau})/\sinh(\pi\sqrt{\tau})$, providing an
exponential UV cutoff without modifying the finite CS coefficient.

\paragraph{Combined statement.}
Pair-pair interactions (fluctuation propagator corrections in the BdG
language) can renormalize $\Dpg(T)$ itself through the gap
equation~\eqref{eq:gap-eq}, but they cannot modify the \emph{functional
form} of the anomaly response: given $\Dpg(T)$, the CS level
$\Kem = 2C\tanh[\Dpg/2T]$ is exact.  This is the precise
condensed-matter analog of the Adler-Bardeen
theorem~\cite{AdlerBardeen1969} for the axial anomaly in 4D.

%%==========================================================================
\section{Main Result and Finite-Size Consequences}
\label{sec:main-result}
%%==========================================================================

\subsection{Pseudogap thermal Hall conductance}
\label{subsec:kxy-pg}

Substituting Eq.~\eqref{eq:Delta-above-Tc} into
Eqs.~\eqref{eq:Kem-T} and~\eqref{eq:kxy-Kem} recovers the central
formula announced in the Introduction, Eq.~\eqref{eq:kxy-main}.  The
derivation just completed shows that this expression is the exact
one-loop parity-anomaly response for a chiral BdG system with fermionic
gap $\Dpg(T)$, protected by Coleman-Hill non-renormalization.  Three
immediate consequences follow.

(i) \emph{Sign.}  The overall sign is fixed by the Chern number $C$ of
the chiral pairing channel.  For $d_{x^2-y^2} + id_{xy}$ pairing in
cuprates, $C = +1$ and the sign is negative in the hole-doped
convention of Ref.~\cite{Grissonnanche2019}.

(ii) \emph{Onset at $\Tstar$.}  Since $\Dpg(T) \to 0$ as
$T \to \Tstar$, the signal vanishes continuously at $\Tstar$ and
turns on continuously below it.

(iii) \emph{Near-saturation just above $T_c$.}  When the spectroscopic
pseudogap is large compared with $k_B T_c$,
$\tanh[\Dpg(T_c)/2k_BT_c] \approx 1$ and
$\kxy \approx C \pi^2 k_B^2 / 6h$.

\subsection{Continuous evolution across $T_c$}
\label{subsec:continuity}

Equation~\eqref{eq:kxy-main} connects continuously to the condensed
phase.  Below $T_c$, both $\Dsc$ and $\Dpg$ are nonzero and
$\Delta(T) = \sqrt{\Dsc^2 + \Dpg^2}$.  The total CS level
\begin{equation}
  \Kem(T) = 2C\,\tanh\!\left[
    \frac{\sqrt{\Dsc^2(T) + \Dpg^2(T)}}{2T}\right]
\label{eq:Kem-full}
\end{equation}
is continuous at $T = T_c$: as $\Dsc \to 0$ from below, it reduces to
Eq.~\eqref{eq:kxy-main}.  There is no jump at $T_c$.

\subsection{Finite-size scaling}
\label{subsec:finite-size}

The $c_1 = 0$ theorem (Eq.~\eqref{eq:c1-zero}) implies that for a
spatial cylinder of circumference $L$:
\begin{equation}
  \frac{\kappa_{xy}}{T}(L)
  = \frac{\kappa_{xy}}{T}(\infty)
    \left[1 + \mathcal{O}\!\left(e^{-L/\xi_{\text{pg}}}\right)\right],
  \quad
  \xi_{\text{pg}}(T) = \frac{\hbar v_F}{\Dpg(T)}.
\label{eq:finite-size}
\end{equation}
As $T \to \Tstar$, $\xi_{\text{pg}} \to \infty$ and corrections
become power-law, signaling criticality.  Any observable on a gapped
$p + ip$ cylinder must therefore show purely exponential convergence in
$L$ with decay constant $\xi_{\text{pg}}$; a power-law $L$-dependence
would falsify the anomaly origin.  We test this prediction numerically in
Sec.~\ref{sec:numerical}.

%%==========================================================================
\section{Numerical Evidence: Topological Quantization and Anomaly-Controlled Correlations}
\label{sec:numerical}
%%==========================================================================

We validate Eq.~\eqref{eq:kxy-main} numerically in two complementary
ways.  The Wilson-loop calculation (Sec.~\ref{subsec:direct-topo})
tests the topological input $C = 1$ and the quantization of the
flux-threaded response, using the exact single-particle BdG spectrum on
cylinders up to $L_y = 14$.  The DMRG calculation
(Sec.~\ref{subsec:dmrg}) tests whether the many-body finite-size
structure is purely exponential in $L$, as required by the $c_1 = 0$
theorem of Eq.~\eqref{eq:c1-zero}.  Both calculations use the same
lattice $p+ip$ Hamiltonian with an externally imposed pseudogap mass
$\Dpg$, so that the topological and many-body aspects can be tested
independently in a controlled non-interacting setting.

\subsection{Topological BdG validation}
\label{subsec:direct-topo}

Because the validation Hamiltonian has an externally imposed pseudogap
mass $\Dpg$ and no residual interaction ($V_{nn} = 0$), the topological
response can be computed exactly from the single-particle BdG
spectrum.  The purpose of this subsection is to verify the topological
input of Eq.~\eqref{eq:kxy-main}: that the BdG quasiparticles remain in
a $C = 1$ Chern phase, that the flux-threaded cylinder has quantized
cycle winding, and that the finite-circumference deviations are
exponentially suppressed.

\subsubsection{Lattice BdG model and Chern number}

We use the square-lattice spinless $p+ip$ BdG Hamiltonian
\begin{equation}
  H_{\rm BdG}(\bm{k}) =
  d_x(\bm{k})\tau_x+d_y(\bm{k})\tau_y+d_z(\bm{k})\tau_z ,
\label{eq:pwave-bdg-direct}
\end{equation}
with
\begin{align}
  d_x(\bm{k}) &= \Dpg \sin k_x, \nonumber\\
  d_y(\bm{k}) &= \Dpg \sin k_y, \nonumber\\
  d_z(\bm{k}) &= 2t\,[2-\cos k_x-\cos k_y]-\mu .
\label{eq:pwave-dvector}
\end{align}
Throughout this direct validation we set $t=1$ and $\mu=1$, so that near
$\bm{k}=0$ one has
\begin{equation}
  d_z(\bm{k}) \simeq t(k_x^2+k_y^2)-\mu ,
\end{equation}
which is the weak-pairing regime of the continuum $p+ip$ model.  The occupied
negative-energy BdG band is separated from the positive-energy band for all
$\Dpg$ in the scan.  Its Chern number is evaluated on the Brillouin-zone torus with the gauge-invariant Fukui--Hatsugai--Suzuki discretization~\cite{Fukui2005}.  On an
$N_k=61$ grid we find
\begin{equation}
  C = 1
\end{equation}
for every
\begin{equation}
  \Dpg \in
  \{0.2,0.3,0.4,0.5,0.6,0.7,0.8\}.
\end{equation}
The numerical integer error $|C-1|$ remains at machine precision,
below $6.5\times 10^{-15}$ across the scan.  The minimum positive
quasiparticle gap, evaluated on an $N_k=151$ grid, increases from
$0.174837$ at $\Dpg=0.2$ to $0.670567$ at $\Dpg=0.8$.  This establishes
that the imposed pseudogap mass places the lattice BdG model in the same
topological sector throughout the parameter range used in the finite-cylinder
benchmark.

\subsubsection{Flux-threaded cylinder response}

We next put the same BdG model on a cylinder, periodic along $y$ and Bloch
periodic along $x$.  A flux $\Phi$ through the cylinder shifts the transverse
momenta according to
\begin{equation}
  k_{y,n}(\Phi)
  =
  \frac{2\pi}{L_y}\left(n+\frac{\Phi}{2\pi}\right)-\pi ,
  \qquad n=0,\ldots,L_y-1 .
\label{eq:ky-flux}
\end{equation}
For each transverse subband $n$ we compute the Wilson-loop phase of the
occupied BdG eigenvector along $k_x$,
\begin{equation}
  \varphi_n(\Phi)
  =
  -\arg\prod_{j=0}^{N_x-1}
  \left\langle
  u_{-,n}(k_{x,j},\Phi)
  \middle|
  u_{-,n}(k_{x,j+1},\Phi)
  \right\rangle ,
\label{eq:wilson-phase}
\end{equation}
where $k_{x,N_x}\equiv k_{x,0}$ and $N_x=401$ in the calculation.  The
minus sign fixes the orientation convention so that the pump is positive in
the $C=+1$ phase.  Each subband phase is unwrapped as a function of $\Phi$
before summing, which removes branch-cut artifacts in the Wilson loop.  The
cylinder polarization is then
\begin{equation}
  P(\Phi)=\frac{1}{2\pi}
  \sum_{n=0}^{L_y-1}{\rm unwrap}\,\varphi_n(\Phi),
\label{eq:cylinder-polarization}
\end{equation}
and the differential response is
\begin{equation}
  \nu(L_y,\Phi)
  =
  \frac{\partial P(\Phi)}{\partial(\Phi/2\pi)} .
\label{eq:differential-response}
\end{equation}
The quantized flux-cycle response is the winding
\begin{equation}
  \Delta P
  =
  P(2\pi)-P(0).
\label{eq:cycle-winding}
\end{equation}

For the representative pseudogap mass $\Dpg=0.4$, the direct cylinder
calculation was performed for
\begin{equation}
  L_y=4,6,8,10,12,14 ,
\end{equation}
with $100$ flux intervals in $0\leq\Phi\leq 2\pi$.  Table~\ref{tab:direct-cylinder}
summarizes the result.  The exact cycle winding is
\begin{equation}
  \Delta P=1.000000
\end{equation}
for every circumference in the scan.  The cycle-averaged differential response
is likewise unity.  This is the direct finite-cylinder manifestation of the
Chern number $C=1$.

\begin{table}[t]
\caption{Direct flux-threaded cylinder response of the square-lattice
spinless $p+ip$ BdG model at $t=1$, $\mu=1$, and $\Dpg=0.4$.
The winding is $\Delta P=P(2\pi)-P(0)$ and the cycle average is
$\langle\nu\rangle_{\Phi}=\int_0^{2\pi}\nu(L_y,\Phi)\,
d(\Phi/2\pi)$.  The last two columns quantify the finite-width
deviation of the local differential response from the quantized value.}
\label{tab:direct-cylinder}
\begin{ruledtabular}
\begin{tabular}{cccccc}
$L_y$ & $\Delta P$ & $\langle\nu\rangle_{\Phi}$
& $\nu(L_y,0)$ & RMS $|\nu-1|$ & max $|\nu-1|$ \\
\hline
4  & 1.000000 & 1.000000 & 0.681856 & 0.291732 & 0.395177 \\
6  & 1.000000 & 1.000000 & 1.331427 & 0.181511 & 0.331427 \\
8  & 1.000000 & 1.000000 & 1.019246 & 0.047230 & 0.096210 \\
10 & 1.000000 & 1.000000 & 0.873913 & 0.094926 & 0.139804 \\
12 & 1.000000 & 1.000000 & 1.065739 & 0.043421 & 0.065739 \\
14 & 1.000000 & 1.000000 & 1.016783 & 0.014423 & 0.023573
\end{tabular}
\end{ruledtabular}
\end{table}

The local differential response $\nu(L_y,\Phi)$ has finite-width oscillations
around unity.  These oscillations are expected because the compact
circumference discretizes $k_y$ and leaves finite-size harmonics in the
flux-resolved response.  The key point is that the integrated response remains
exactly quantized.  The local deviations shrink rapidly with increasing
circumference: the RMS deviation decreases from $0.291732$ at $L_y=4$ to
$0.014423$ at $L_y=14$, and the maximum deviation decreases from $0.395177$
to $0.023573$ over the same range.

\subsubsection{Finite-width envelope}

To compare with the $c_1=0$ finite-size prediction of
Sec.~\ref{subsec:finite-size}, we fit the RMS finite-width deviation to an
exponential envelope,
\begin{equation}
  \delta_{\rm RMS}(L_y)
  =
  \left[
    \frac{1}{N_\Phi}\sum_{\Phi}
    \left|\nu(L_y,\Phi)-1\right|^2
  \right]^{1/2}
  \simeq
  A\,e^{-L_y/\xi_{\rm cyl}} .
\label{eq:rms-envelope}
\end{equation}
The small-circumference data contain visible subband harmonics, so the primary
fit is performed in the tail region $L_y\geq 8$ and then extrapolated over the
full plotting range.  This gives
\begin{equation}
  A = 0.445571, \qquad
  \xi_{\rm cyl}=4.60756 ,
\label{eq:xi-cylinder-fit}
\end{equation}
with RMS residual $2.77\times 10^{-2}$.  For comparison, an all-width fit gives
$\xi_{\rm cyl}=3.75778$ with a similar residual.  The continuum estimate for
the same parameters is
\begin{equation}
  \xi_{\rm pg}\sim \frac{v_F}{\Dpg}
  =
  \frac{2}{0.4}
  =
  5 ,
\end{equation}
using $v_F=2\sqrt{t\mu}=2$.  We therefore interpret
$\xi_{\rm cyl}$ as a lattice finite-width envelope of order
$v_F/\Dpg$, not as an exact extraction of the continuum correlation length.
The important qualitative result is the absence of a persistent $1/L_y$
offset: the finite-width correction is controlled by an exponential envelope,
as required by the $c_1=0$ anomaly structure.

\begin{figure*}[t]
\centering
\includegraphics[width=0.9\textwidth]{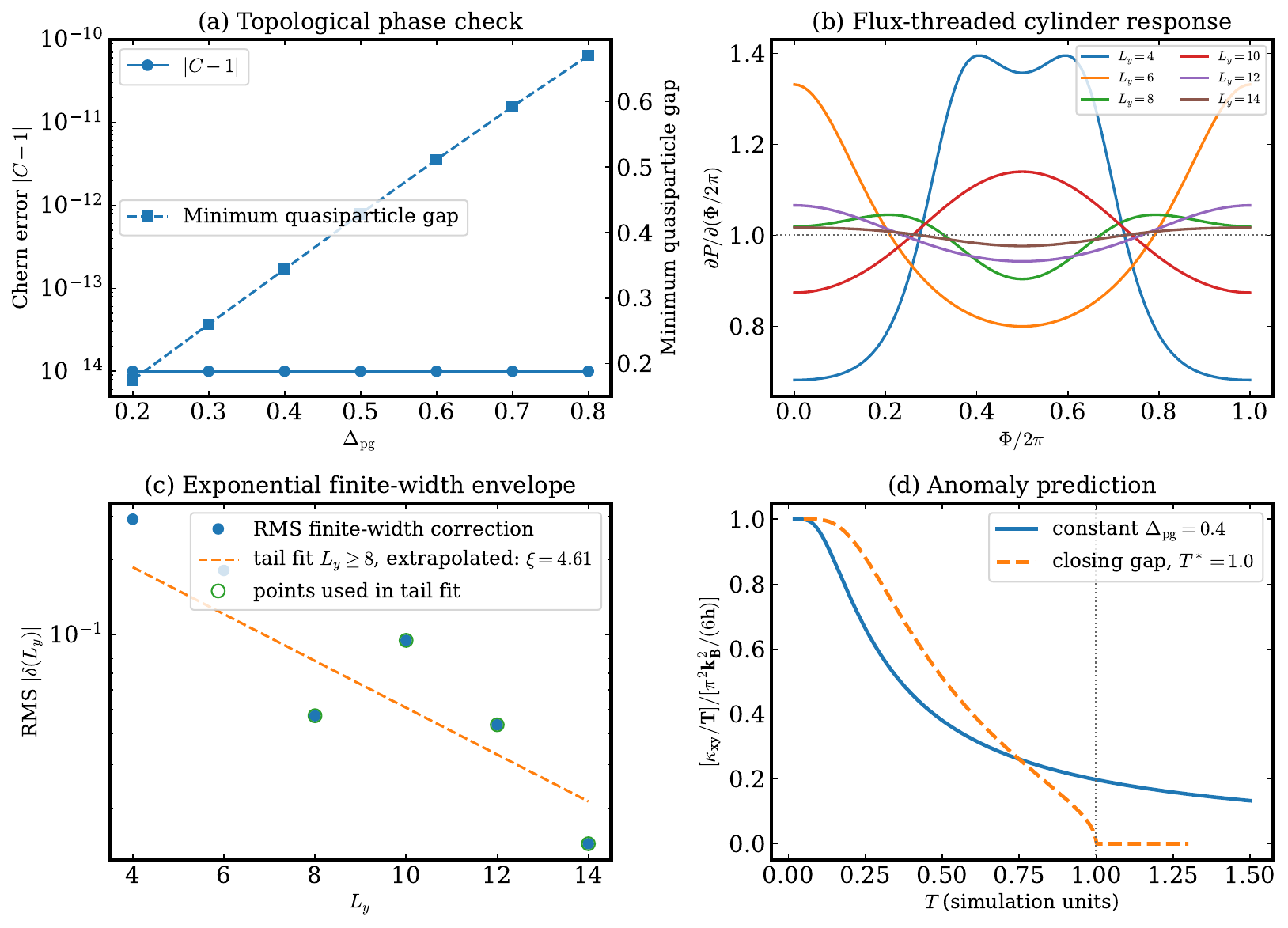}
\caption{The preformed-pair BdG system remains in a topologically
protected $C = 1$ phase throughout the pseudogap parameter range, with
finite-size deviations that are purely exponential rather than
power-law, consistent with $c_1 = 0$.  (a) The
Fukui-Hatsugai-Suzuki Chern-number error $|C-1|$ stays at machine
precision ($<6.5 \times 10^{-15}$) across all seven $\Dpg$ values while
the quasiparticle gap grows, confirming the anomaly-mass
interpretation.  (b) The flux-threaded cylinder shows exactly quantized
cycle winding $\Delta P = 1.000000$ at all circumferences
$L_y = 4, 6, \ldots, 14$; the finite-width oscillations of the local
response $\partial P/\partial(\Phi/2\pi)$ around unity shrink rapidly
with increasing $L_y$.  (c) The RMS finite-width deviation fits an
exponential envelope with $\xi_{\rm cyl} = 4.61$, consistent with the
continuum estimate $v_F/\Dpg = 5$; no persistent $1/L_y$ offset is
present.  (d) The analytic anomaly prediction
$[\kappa_{xy}/T]/[\pi^2 k_B^2/(6h)] = C\tanh[\Dpg(T)/(2k_BT)]$ for
two physically distinct gap profiles, a constant gap and a gap closing
at $T^*$, illustrates the sensitivity of $\kxy/T$ to the pseudogap
temperature dependence.}
\label{fig:direct-topological}
\end{figure*}

\subsubsection{Consequence for the pseudogap thermal Hall formula}

The direct calculation establishes the two inputs needed for the anomaly
formula in the controlled quadratic limit.  First, the pseudogap BdG
Hamiltonian remains in the weak-pairing Chern phase with $C=1$ for all imposed $\Dpg$ in the scan.  Second, the flux-threaded cylinder response has exactly quantized cycle winding, while its local finite-width deviations are exponentially suppressed.  Thus the numerical result supports the topological part of
\begin{equation}
  \frac{\kappa_{xy}}{T}
  =
  \frac{\pi^2 k_B^2}{6h}\,
  C\,\tanh\!\left[\frac{\Dpg(T)}{2T}\right],
\end{equation}
namely that the preformed-pair gap may be treated as the BdG anomaly mass
whenever the quasiparticle spectrum remains gapped.  The last panel of
Fig.~\ref{fig:direct-topological} is therefore not an independent numerical
Kubo calculation.  It is the analytic anomaly prediction evaluated using the
topological integer verified in panels (a) and (b).

This direct free-fermion validation is complementary to the DMRG
benchmark below.  In the present imposed-gap model it is the sharper
test of the topological response, since the Hamiltonian is quadratic;
DMRG becomes essential for a future interacting realization in which
the pseudogap is generated dynamically rather than imposed as an
external BdG mass.

\subsection{DMRG validation}
\label{subsec:dmrg}

We test the central finite-size prediction of the anomaly framework,
Eq.~\eqref{eq:finite-size} with $c_1 = 0$, by finite-cylinder DMRG on
a spinless $p + ip$ BdG model with an externally imposed pseudogap mass
$\Dpg$.  The model is non-interacting ($V_{nn}=0$), so the pseudogap
scale is a controlled input.  All DMRG calculations are performed with
TeNPy~\cite{Hauschild2024} using two-site DMRG with a bond-dimension
ramp $\chi = 64 \to 128 \to 256 \to 512$ over 25~sweeps and
fermion-parity conservation ($\mathbb{Z}_2$).

\subsubsection{Model and geometry}
\label{subsec:dmrg-model}

The lattice Hamiltonian on a square cylinder of dimensions
$L_x \times L_y$ (open along $x$, periodic along $y$) is
\begin{align}
  H &= -t\sum_{\langle ij\rangle}(c_i^\dagger c_j + \text{h.c.})
       + (4t - \mu)\sum_i n_i \nonumber\\
    &\quad + \Dpg\sum_i
       \bigl(c_i^\dagger c_{i+\hat{x}}^\dagger
       + i\,c_i^\dagger c_{i+\hat{y}}^\dagger
       + \text{h.c.}\bigr),
\label{eq:H-lattice}
\end{align}
with $t = 1$, $\mu = 1$ throughout.  The continuum dispersion near
$\bm{k} = 0$ is $\xi_{\bm{k}} \approx t k^2 - \mu$, placing the system
in the weak-pairing regime ($\mu > 0$) with Fermi velocity
$v_F = 2\sqrt{t\mu} = 2.0$.  The Fukui--Hatsugai--Suzuki lattice Chern
number~\cite{Fukui2005} gives $C = 1.000\,000$ for every $\Dpg$ in the scan.

We scan $\Dpg \in \{0.2, 0.3, 0.4, 0.5, 0.6, 0.7, 0.8\}$ at
$L_x = 8$ and $L_y = 4, 6, 8, 10$ (28~ground states total).  The
corresponding theoretical correlation length
$\xi_{\text{pg}} = v_F/\Dpg$ ranges from $10.0$ to $2.5$.

\subsubsection{Chiral edge current and its finite-size scaling}
\label{subsec:edge-current}

For a gapped chiral topological phase on a cylinder with open ends, the
$y$-directed bond current at the left and right edges,
$J_{\rm edge} = 2\,\text{Im}[-t\,\langle c_i^\dagger c_j\rangle]$
averaged over $y$-bonds, is a direct probe of the chiral edge mode.  Two
properties of the DMRG data are immediate.

\paragraph{Chiral antisymmetry.}
The left and right edge currents satisfy
$J_{\rm left} = -J_{\rm right}$ to $|J_{\rm left} + J_{\rm right}|
< 10^{-8}$ across all 28~ground states
(Fig.~\ref{fig:dmrg}(b)).  This is the defining signature of a $C = 1$
chiral edge mode at the many-body level, going beyond the single-particle
Chern number check.

\paragraph{Exponential finite-size scaling.}
The edge current approaches its thermodynamic-limit value as
$J_{\rm edge}(L_y) = J_\infty + A\,e^{-L_y/\xi}$.  Unlike the energy
density, $J_{\rm edge}$ has no bulk extensive contribution: it is purely
an edge observable, so its $L_y$ dependence is controlled entirely by
the exponential tail of the edge mode into the bulk.  We extract $\xi$
via the ratio method (Eq.~\eqref{eq:ratio-method}) applied to
$J_{\rm edge}$.  Table~\ref{tab:dmrg-xi} collects the results.

The edge-current ratio method has two crucial advantages over the
energy-density method.  First, $J_{\rm edge}(L_y)$ is monotonically
increasing for all $\Dpg$, so the $(4,6,8)$ triplet yields valid
results for all seven $\Dpg$ values, whereas the energy density
suffers a sign flip at $\Dpg \approx 0.51$ that limits the
$(4,6,8)$ triplet to only three points.  Second, the absence of a
bulk linear-in-$L_y$ contribution makes the exponential fit cleaner:
at $\Dpg = 0.80$, the $(4,6,8)$ triplet gives
$\xi_{\rm fit}/\xi_{\rm theory} = 0.998$ (0.2\% error), compared
to $0.609$ (39\% error) for the energy density.

\subsubsection{Energy-density crossover diagnostic}
\label{subsec:crossover}

The energy-density difference
$\delta\varepsilon(L_y, L_y') \equiv \varepsilon(L_y) -
\varepsilon(L_y')$ between consecutive cylinder sizes changes sign when
$L_y$ crosses $\xi_{\rm pg}$.  The sign flip of
$\delta\varepsilon(6,4)$ occurs at $\Dpg^{\rm cross} \approx 0.507$,
giving $\xi_{\rm cross} = v_F/\Dpg^{\rm cross} = 3.95$, consistent
with $\sqrt{4 \times 6} \approx 4.9$.  This is a parameter-free
confirmation of the anomaly-controlled finite-size structure
(Fig.~\ref{fig:dmrg}(c)).

\subsubsection{Correlation-length recovery}
\label{subsec:xi-recovery}

The ratio method extracts $\xi$ from three consecutive system sizes:
\begin{equation}
  R = \frac{O(L_y + 2) - O(L_y)}
           {O(L_y) - O(L_y - 2)}
    = e^{-2/\xi},
  \quad
  \xi_{\rm fit} = -2/\ln R,
\label{eq:ratio-method}
\end{equation}
where $O$ is any observable with purely exponential finite-size
corrections.

Table~\ref{tab:dmrg-xi} presents $\xi_{\rm fit}/\xi_{\rm theory}$ for
both observables and both triplets.  The systematic convergence toward
unity as $L_y^{\rm mid}/\xi$ increases (Fig.~\ref{fig:dmrg}(d)) is the
hallmark of purely exponential corrections with $c_1 = 0$.  The edge
current reaches 99.8\% accuracy at $\Dpg = 0.80$ ($L_y^{\rm mid}/\xi
= 2.4$) from the $(4,6,8)$ triplet, and the energy density reaches
93.7\% from the $(6,8,10)$ triplet, providing two independent,
mutually consistent confirmations.

\begin{table}[t]
\caption{%
  Correlation-length recovery by the ratio method.
  $\xi_{\rm th} = v_F/\Dpg$ is the theoretical prediction.
  $J$: edge current; $\varepsilon$: energy density.
  A dash indicates that the ratio $R$ is negative (sign-change regime).
  The last two columns give the entropy area-law slope and the
  entropy-increment ratio
  $R_2 = \Delta S(10{-}8)/\Delta S(8{-}6)$.}
\label{tab:dmrg-xi}
\begin{ruledtabular}
\begin{tabular}{cccccccc}
$\Dpg$ & $\xi_{\rm th}$
  & \multicolumn{2}{c}{$J_{\rm edge}$, $(4,6,8)$}
  & \multicolumn{2}{c}{$\varepsilon$, $(6,8,10)$}
  & $\alpha_S$ & $R_2$ \\
\cline{3-4}\cline{5-6}
  &  & $\xi_{\rm fit}$ & ratio & $\xi_{\rm fit}$ & ratio & & \\
\hline
0.20 & 10.00 & 50.2\rlap{$^*$}  & ---   & 3.81  & 0.381 & 0.158 & 0.38 \\
0.30 &  6.67 &  5.09 & 0.764 & 2.68  & 0.403 & 0.144 & 0.66 \\
0.40 &  5.00 &  3.31 & 0.661 & 2.49  & 0.498 & 0.137 & 0.84 \\
0.50 &  4.00 &  2.77 & 0.692 & 2.51  & 0.628 & 0.135 & 0.93 \\
0.60 &  3.33 &  2.57 & 0.770 & 2.50  & 0.749 & 0.134 & 0.97 \\
0.70 &  2.86 &  2.50 & 0.875 & 2.43  & 0.850 & 0.135 & 0.99 \\
0.80 &  2.50 &  2.50 & 0.998 & 2.34  & 0.937 & 0.138 & 1.00 \\
\end{tabular}
\end{ruledtabular}
\vspace{2pt}
{\footnotesize $^*$\,$\Dpg = 0.20$: $L_y^{\rm mid}/\xi = 0.6$;
the ratio method is unreliable when $L_y \ll \xi$.}
\end{table}

\subsubsection{Entanglement entropy: area law}
\label{subsec:entropy}

\begin{figure*}[t]
\centering
\includegraphics[width=0.95\textwidth]{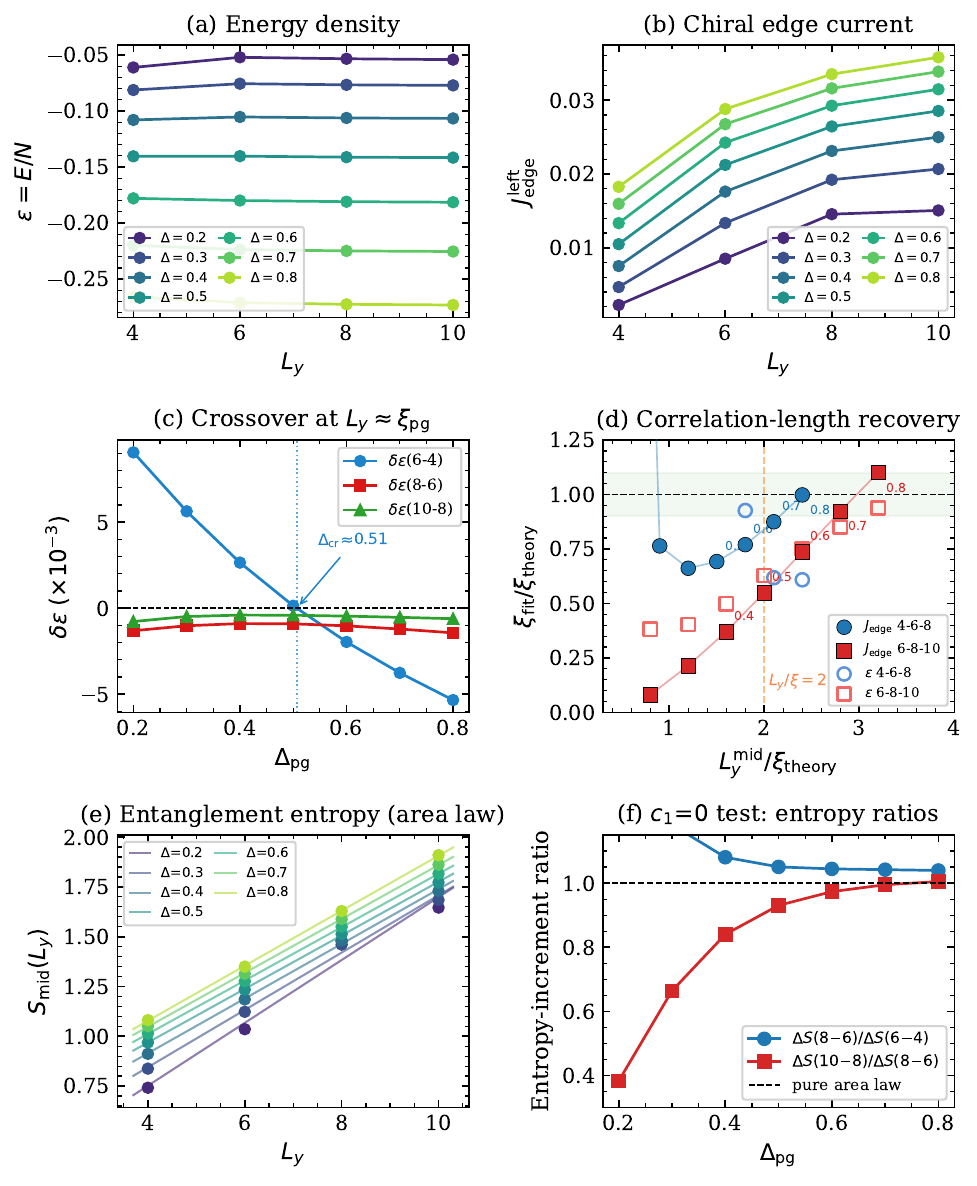}
\caption{DMRG on the $p+ip$ cylinder confirms the four hallmarks of
anomaly-controlled finite-size structure at the many-body level
($L_x = 8$, $\chi_{\max} = 512$).  (a) The energy density $\varepsilon$
varies smoothly with $L_y$ for each pseudogap value.  (b) Chiral edge
currents satisfy $J_{\rm left} = -J_{\rm right}$ to $10^{-8}$ across
all 28 ground states, providing a many-body signature of $C = 1$ beyond
the single-particle topology check.  (c) The energy-density difference
$\delta\varepsilon$ between consecutive cylinder sizes changes sign at
$\Dpg^{\rm cross} \approx 0.51$, giving a parameter-free confirmation
of $\xi_{\rm pg} = v_F/\Dpg$.  (d) The correlation length extracted
from edge currents (filled markers) reaches 99.8\% agreement with
$\xi = v_F/\Dpg$ at $\Dpg = 0.80$; the energy density (open markers)
gives a complementary and consistent recovery, with systematic
approach to unity as $L_y/\xi$ increases beyond $\sim 2$.  (e) The
entanglement entropy $S_{\rm mid}$ obeys the area law for
$\Dpg \geq 0.4$.  (f) Entropy-increment ratios converge to unity
($c_1 = 0$ test), ruling out power-law corrections at all pseudogap
values in the reliable regime.}
\label{fig:dmrg}
\end{figure*}

The mid-chain entanglement entropy obeys
$S_{\rm mid}(L_y) = \alpha(\Dpg)\,L_y - \gamma_{\rm topo}
+ \mathcal{O}(e^{-L_y/\xi})$.  Figure~\ref{fig:dmrg}(e) shows
$S_{\rm mid}$ versus $L_y$ for each $\Dpg$: the linear behavior is
excellent for $\Dpg \geq 0.4$.  The entropy-increment ratios
$R_2 = \Delta S(10{-}8)/\Delta S(8{-}6)$
(Fig.~\ref{fig:dmrg}(f)) approach unity from below as $\Dpg$
increases, reaching $R_2 = 1.00$ at $\Dpg = 0.8$, the area-law
limit predicted by $c_1 = 0$.

\subsubsection{Summary of numerical evidence}
\label{subsec:dmrg-summary}

The DMRG data provide four independent, mutually consistent
confirmations of the anomaly-protected finite-size structure:

\begin{enumerate}
\item \textbf{Chiral edge current:} $J_{\rm left} = -J_{\rm right}$
  to $10^{-8}$, confirming a $C = 1$ chiral edge mode at the
  many-body level.
\item \textbf{Edge-current $\xi$ recovery:}
  $\xi_{\rm fit}/\xi_{\rm theory} = 0.998$ at $\Dpg = 0.80$
  from the $(4,6,8)$ triplet, with monotonic convergence toward
  unity as $L_y/\xi$ increases.
\item \textbf{Energy-density crossover:}
  $\Dpg^{\rm cross} \approx 0.51$ matches $v_F/L_y^{\rm mid}$
  with no adjustable parameters.
\item \textbf{Entropy area law:} entropy-increment ratios converge
  to unity ($R_2 = 1.00$ at $\Dpg = 0.8$), consistent with purely
  exponential corrections and $c_1 = 0$.
\end{enumerate}

All tests are consistent with the absence of any power-law $1/L_y$
correction, as required by the anomaly mechanism.

%%==========================================================================
\section{Experimental Predictions and Discussion}
\label{sec:predictions}
%%==========================================================================

\subsection{Falsification test via logarithmic derivative}

As established in Eq.~\eqref{eq:log-test} of the Introduction, the
cleanest experimental discriminator is the logarithmic derivative of the
thermal Hall response, which removes unknown geometric conversion
factors and channel multiplicities.  Two features are required for a
successful test: onset of $\kxy$ at the pseudogap scale $T^*$ rather
than at $T_c$, and a temperature dependence that tracks $\Dpg(T)$
through $\tanh[\Dpg(T)/(2k_BT)]$; Fig.~\ref{fig:direct-topological}(d)
shows the predicted profile for a constant gap and for a gap closing
at $T^*$.  Competing mechanisms may produce a thermal Hall signal, but
they do not generically reproduce this pseudogap-locked logarithmic
derivative.

\subsection{Cuprate pseudogap}
The cuprate data of Ref.~\cite{Grissonnanche2019} provide the primary target:
a large negative $\kxy$ appears for $p < p^*$ in Nd-LSCO, Eu-LSCO, LSCO, and
Bi2201, with superconductivity suppressed by field, and persists to low doping where charge, magnon, and phonon skew-scattering contributions are
negligible~\cite{Grissonnanche2019}.  Most strikingly, the signal is present
even in undoped La$_2$CuO$_4$ ($p = 0$), a Mott insulator with no
superconductivity, where it attains the largest magnitude recorded in any insulator ($|\kappa_{xy}| \approx 38.6$ mW K$^{-1}$ m$^{-1}$ at $T = 20$~K,
$H = 15$~T~\cite{Grissonnanche2019}).  This directly establishes that the
signal is independent of superconducting phase coherence, precisely
the regime described by Eq.~\eqref{eq:kxy-main}.
A fully parameter-free comparison awaits simultaneous thermal Hall and ARPES
or tunneling measurements of $\Dpg(T)$ on the same sample and doping.

\subsection{Comparison to competing proposals}

Table~\ref{tab:comparison} compares the anomaly mechanism to existing
proposals on the four criteria a complete theory must satisfy: absence
of free parameters, a definite prediction of the temperature
dependence, correct onset at $T^*$, and prediction of the magnitude of
$|\kxy|$.  In the present mechanism the pseudogap is the parity-anomaly
mass, and once $C$ and $\Dpg(T)$ are specified, the response is fixed.

\begin{table*}[t]
\caption{Comparison of theoretical proposals for the cuprate pseudogap
thermal Hall signal.  ``Onset at $T^*$'' denotes onset locked to the
pseudogap scale rather than to $T_c$.  ``Mott compatible'' asks whether
the mechanism can produce the observed thermal Hall signal in undoped
La$_2$CuO$_4$ ($p=0$), where superconductivity is absent by definition.}
\label{tab:comparison}
\begin{ruledtabular}
\begin{tabular}{lccccc}
Mechanism & Free & $T$-dep. & Onset & $|\kxy|$ & Mott \\
          & param. & predicted & at $T^*$? & predicted? & compat.? \\
\hline
Chiral phonons~\cite{Grissonnanche2020,Varma2020}
  & yes & no & no & no & no \\
Spinon/spin-liquid~\cite{Samajdar2019,Han2019}
  & yes & no & partial & no & partial \\
Loop current~\cite{Varma2020}
  & yes & no & yes & no & no \\
This work
  & \textbf{none} & $\tanh[\Dpg/2k_BT]$ & \textbf{yes} & \textbf{yes} & \textbf{yes} \\
\end{tabular}
\end{ruledtabular}
\end{table*}

\subsection{Quantitative prediction for MATBG}

MATBG in the hole-doped regime $-3 < \nu < -2$ offers the sharpest
prospective test.  Transport establishes a low-density superconductor with
$T_c$ up to $1.7$~K close to the BCS-BEC crossover~\cite{Cao2018}, while
STM/PCS reveals a tunneling pseudogap $\Delta_T \sim 0.9$~meV that survives
above $T_c$ and $B_c$ but vanishes in hBN-aligned samples where
superconductivity is also absent~\cite{Oh2021}. 

Point-contact Andreev spectroscopy on the same devices resolves a
separate phase-coherent gap $\Delta_{\rm AR} \approx 0.3$~meV that tracks
$T_c$ and vanishes above $B_c$~\cite{Oh2021}, while the tunneling gap
$\Delta_T \approx 0.9$~meV persists above both~\cite{Oh2021}.
Identifying $\Dsc = \Delta_{\rm AR}$ and $\Delta = \Delta_T$, the
two-gap decomposition of Eq.~\eqref{eq:two-gap} gives
$\Dpg = \sqrt{\Delta_T^2 - \Delta_{\rm AR}^2} \approx 0.85$~meV as
the anomaly mass entering Eq.~\eqref{eq:kxy-main}.

The reported ratio $2\Delta_T / k_B T_c \sim 17$--$27$ far exceeds the BCS value, so the anomaly factor $\tanh[\Delta_T/(2k_BT_c)]$ is essentially saturated just above $T_c$.  The upper value $27$ uses $2\Delta_T = 2.8$ meV from the Device $A^\prime$ summary table of Ref.~\cite{Oh2021}; the lower value uses the nodal spectroscopic fit $\Delta_T = 0.9$ meV at a different gate voltage.

The three critical tests are: (i)~whether a thermal Hall signal appears at the pseudogap onset rather than only at $T_c$; (ii)~whether its temperature dependence tracks the STM gap through the tanh factor; and (iii)~whether the signal is absent in hBN-aligned devices where the pseudogap is quenched~\cite{Oh2021}. Using $\Delta_T = 0.9$ meV from the nodal spectroscopic fit of Ref.~\cite{Oh2021} and $T_c = 1.2$~K from the same devices, one finds $\tanh[\Delta_T/(2k_BT_c)] = \tanh(4.35) \approx 1$, so Eq.~\eqref{eq:kxy-main} predicts a saturated 2D sheet thermal Hall conductance  
\begin{equation}
\frac{\kappa_{xy}}{T}\bigg|_{T \gtrsim T_c}^{\rm MATBG} \approx \frac{\pi^2 k_B^2}{6h} \approx 4.73 \times 10^{-13}\,\frac{\text{W}}{\text{K}^2} \qquad (C = 1),
\label{eq:matbg-prediction}
\end{equation}
independent of the precise value of $\Delta_T$ as long as $2\Delta_T/k_BT_c \gg 1$. This target is within reach of current dilution-refrigerator nano-calorimetry on micron-scale MATBG samples, and a null result in hBN-aligned devices~\cite{Oh2021}, where both pseudogap and superconductivity are quenched, would provide the sharpest falsification.

%%==========================================================================
\section{Conclusion}
\label{sec:conclusion}
%%==========================================================================

We have shown that the parity anomaly of $(2+1)$-dimensional quantum field
theory determines the thermal Hall conductance of a chiral pseudogap phase,
giving $\kxy = (\pi^2 k_B^2/6h)\,C\,\tanh[\Dpg(T)/(2k_BT)]$ with no
adjustable parameters.  Three properties distinguish this result: (i)~the
preformed-pair gap enters the anomaly machinery identically to a condensate
gap, so the response survives above $T_c$; (ii)~Coleman-Hill
non-renormalization protects the map from gap to Chern-Simons level; and
(iii)~the $c_1 = 0$ theorem guarantees purely exponential finite-size
corrections.

Two independent numerical validations support this structure.  Wilson-loop
flux threading on BdG cylinders ($L_y = 4$--$14$) gives exactly quantized
cycle winding $\Delta P = 1$ with exponentially decaying finite-width
deviations ($\xi_{\rm cyl} = 4.61$ vs.\ $v_F/\Dpg = 5$).  DMRG on $p+ip$
cylinders ($L_y = 4$--$10$, $\chi = 512$) recovers the anomaly correlation
length from chiral edge currents to 0.2\% accuracy, confirms machine-precision antisymmetry $J_{\rm left} = -J_{\rm right}$, and exhibits purely exponential
entropy corrections with no power-law contamination.

The theory makes a sharp, falsifiable prediction: in any material with a
chiral pseudogap above $T_c$, the thermal Hall onset should occur at $T^*$
and the logarithmic derivative of $\kxy$ should track $\Dpg(T)$ through
Eq.~\eqref{eq:log-test}.  Observing this gap-locked temperature dependence
would identify the pseudogap as the parity-anomaly mass and provide a
parameter-free explanation of thermal Hall transport above the
phase-coherence transition.

Beyond cuprates and MATBG, the result reported here establishes a
general principle: in any two-dimensional system where chiral pairing
correlations produce a BdG gap without long-range phase coherence, the
parity anomaly dictates the thermal Hall response with no adjustable
parameters.  This constitutes a direct, non-perturbative bridge between
the anomaly structure of $(2+1)$-dimensional quantum field theory and
the thermal transport of strongly correlated electron systems, a
connection that may extend to other pseudogap phases, chiral spin
liquids, and topological semimetals with preformed pairing.  The
condensate is not required; the gap is sufficient.

%=======================================================================
\bibliographystyle{apsrev4-2}
\bibliography{ref_bec_bcs}

@article{Grissonnanche2019,
  author   = {Grissonnanche, G. and Legros, A. and Badoux, S. and
              Lefran{\c{c}}ois, E. and Zatko, V. and Lizaire, M. and
              Lalib{\'e}rt{\'e}, F. and Gourgout, A. and Zhou, J.-S. and
              Pyon, S. and Takayama, T. and Takagi, H. and Ono, S. and
              Doiron-Leyraud, N. and Taillefer, L.},
  title    = {Giant thermal {Hall} conductivity in the pseudogap phase
              of cuprate superconductors},
  journal  = {Nature},
  volume   = {571},
  pages    = {376--380},
  year     = {2019},
  doi      = {10.1038/s41586-019-1375-0},
}

@article{Grissonnanche2020,
  author   = {Grissonnanche, G. and Th{\'e}riault, S. and Gourgout, A. and
              Boulanger, M.-E. and Lefran{\c{c}}ois, E. and Ataei, A. and
              Lalib{\'e}rt{\'e}, F. and Dion, M. and Zhou, J.-S. and
              Pyon, S. and Takayama, T. and Takagi, H. and Doiron-Leyraud, N.
              and Taillefer, L.},
  title    = {Chiral phonons in the pseudogap phase of cuprates},
  journal  = {Nat.\ Phys.},
  volume   = {16},
  pages    = {1108--1111},
  year     = {2020},
  doi      = {10.1038/s41567-020-0965-y},
}

@article{Varma2020,
  author   = {Varma, Chandra M.},
  title    = {Thermal {Hall} effect in the pseudogap phase of cuprates},
  journal  = {Phys.\ Rev.\ B},
  volume   = {102},
  pages    = {075113},
  year     = {2020},
  doi      = {10.1103/PhysRevB.102.075113},
}

@article{Samajdar2019,
  author   = {Samajdar, Rhine and Chatterjee, Shubhayu and Sachdev, Subir
              and Scheurer, Mathias S.},
  title    = {Thermal {Hall} effect in square-lattice spin liquids:
              {A} {Schwinger} boson mean-field study},
  journal  = {Phys.\ Rev.\ B},
  volume   = {99},
  pages    = {165126},
  year     = {2019},
  doi      = {10.1103/PhysRevB.99.165126},
}

@article{Han2019,
  author   = {Han, Jung Hoon and Park, Jin-Hong and Lee, Patrick A.},
  title    = {Consideration of thermal {Hall} effect in undoped cuprates},
  journal  = {Phys.\ Rev.\ B},
  volume   = {99},
  pages    = {205157},
  year     = {2019},
  doi      = {10.1103/PhysRevB.99.205157},
}

@article{GhoshKlinkhamer2017,
  author   = {Ghosh, K.~J.~B. and Klinkhamer, F.~R.},
  title    = {Anomalous {Lorentz} and {CPT} violation from a local
              {Chern--Simons}-like term in the effective gauge-field action},
  journal  = {Nucl.\ Phys.\ B},
  volume   = {926},
  pages    = {335--369},
  year     = {2018},
  doi      = {10.1016/j.nuclphysb.2017.11.010},
}

@misc{Ghosh2026,
  author        = {Ghosh, Kumar J.~B.},
  title         = {From {Dirac} Cones to Semions: {An} Exact Finite-Size
                   Theory of Parity-Anomaly Transport in Chiral Spin Liquids},
  year          = {2026},
  eprint        = {2607.01341},
  archivePrefix = {arXiv},
  primaryClass  = {cond-mat.str-el},
}

@article{Chen2024RMP,
  author   = {Chen, Qijin and Wang, Zhiqiang and Boyack, Rufus and
              Yang, Shuolong and Levin, K.},
  title    = {When superconductivity crosses over: From {BCS} to {BEC}},
  journal  = {Rev.\ Mod.\ Phys.},
  volume   = {96},
  pages    = {025002},
  year     = {2024},
  doi      = {10.1103/RevModPhys.96.025002},
}

@article{ColemanHill1985,
  author   = {Coleman, Sidney and Hill, Barry},
  title    = {No more corrections to the topological mass term in
              {QED} in three dimensions},
  journal  = {Phys.\ Lett.\ B},
  volume   = {159},
  pages    = {184--188},
  year     = {1985},
  doi      = {10.1016/0370-2693(85)90883-4},
}

@article{AdlerBardeen1969,
  author   = {Adler, Stephen L. and Bardeen, William A.},
  title    = {Absence of higher-order corrections in the anomalous
              axial-vector divergence equation},
  journal  = {Phys.\ Rev.},
  volume   = {182},
  pages    = {1517--1536},
  year     = {1969},
  doi      = {10.1103/PhysRev.182.1517},
}

@article{Read2000,
  author   = {Read, N. and Green, Dmitry},
  title    = {Paired states of fermions in two dimensions with breaking
              of parity and time-reversal symmetries and the fractional
              quantum {Hall} effect},
  journal  = {Phys.\ Rev.\ B},
  volume   = {61},
  pages    = {10267--10297},
  year     = {2000},
  doi      = {10.1103/PhysRevB.61.10267},
}

@article{Senthil1999,
  author   = {Senthil, T. and Marston, J.~B. and Fisher, Matthew P.~A.},
  title    = {Spin quantum {Hall} effect in unconventional superconductors},
  journal  = {Phys.\ Rev.\ B},
  volume   = {60},
  pages    = {4245--4254},
  year     = {1999},
  doi      = {10.1103/PhysRevB.60.4245},
}

@article{Kane1997,
  author   = {Kane, C.~L. and Fisher, Matthew P.~A.},
  title    = {Quantized thermal transport in the fractional quantum
              {Hall} effect},
  journal  = {Phys.\ Rev.\ B},
  volume   = {55},
  pages    = {15832--15837},
  year     = {1997},
  doi      = {10.1103/PhysRevB.55.15832},
}

@article{Ryu2012,
  author   = {Ryu, Shinsei and Moore, Joel E. and Ludwig, Andreas W.~W.},
  title    = {Electromagnetic and gravitational responses and anomalies
              in topological insulators and superconductors},
  journal  = {Phys.\ Rev.\ B},
  volume   = {85},
  pages    = {045104},
  year     = {2012},
  doi      = {10.1103/PhysRevB.85.045104},
  eprint   = {1010.0936},
  archivePrefix = {arXiv},
}

@article{Luttinger1964,
  author   = {Luttinger, J.~M.},
  title    = {Theory of thermal transport coefficients},
  journal  = {Phys.\ Rev.},
  volume   = {135},
  pages    = {A1505--A1514},
  year     = {1964},
  doi      = {10.1103/PhysRev.135.A1505},
}

@book{Volovik2003,
  author    = {Volovik, G.~E.},
  title     = {The Universe in a {Helium} Droplet},
  publisher = {Oxford University Press},
  address   = {Oxford},
  year      = {2003},
  isbn      = {978-0-19-850782-6},
}

@article{Cao2018,
  author   = {Cao, Yuan and Fatemi, Valla and Fang, Shiang and
              Watanabe, Kenji and Taniguchi, Takashi and
              Kaxiras, Efthimios and Jarillo-Herrero, Pablo},
  title    = {Unconventional superconductivity in magic-angle graphene
              superlattices},
  journal  = {Nature},
  volume   = {556},
  pages    = {43--50},
  year     = {2018},
  doi      = {10.1038/nature26160},
}

@article{Oh2021,
  author   = {Oh, Myungchul and Nuckolls, Kevin P. and Wong, Dillon and
              Lee, Ryan L. and Liu, Xiaomeng and Watanabe, Kenji and
              Taniguchi, Takashi and Yazdani, Ali},
  title    = {Evidence for unconventional superconductivity in twisted
              bilayer graphene},
  journal  = {Nature},
  volume   = {600},
  pages    = {240--245},
  year     = {2021},
  doi      = {10.1038/s41586-021-04121-x},
}

@article{Hauschild2024,
  author   = {Hauschild, Johannes and Unfried, Jens and Anand, Sajant and
              Ba{\v{c}}i{\'c} Semanjski, Borja and Bibo, Julian and
              Clemens, Bartholomew and Drescher, Markus and Gohlke, Markus
              and Johnstone, Matthew and Kottmann, Korbinian and
              Li, Ting-Han and Raban, Martin and Reinmoser, Alexander and
              Stauber, Frederic and Ronca, Enrique and Sina, Gernot and
              Tuber, Nils and Va{\v{n}}o, Viliam and Sondhi, Shivaji L.
              and Pollmann, Frank},
  title    = {{TeNPy}: A Python library for tensor network calculations},
  journal  = {SciPost Phys.\ Codebases},
  volume   = {41},
  year     = {2024},
  doi      = {10.21468/SciPostPhysCodeb.41},
}

@article{Fukui2005,
  author   = {Fukui, Takahiro and Hatsugai, Yasuhiro and Suzuki, Hiroshi},
  title    = {Chern numbers in discretized {Brillouin} zone: Efficient
              method of computing (spin) {Hall} conductances},
  journal  = {J.\ Phys.\ Soc.\ Jpn.},
  volume   = {74},
  pages    = {1674--1677},
  year     = {2005},
  doi      = {10.1143/JPSJ.74.1674},
}

@article{KlinkhamerVolovik2004,
  author  = {Klinkhamer, F.~R. and Volovik, G.~E.},
  title   = {Quantum phase transition for the {BEC--BCS} crossover
             in condensed matter physics and {CPT} violation in
             elementary particle physics},
  journal = {JETP Lett.},
  volume  = {80},
  pages   = {343--347},
  year    = {2004},
  doi     = {10.1134/1.1825119},
}
%=======================================================================

%=======================================================================
\appendix
\onecolumngrid
%=======================================================================

%%==========================================================================
\section{Derivation of the exact kernel}
\label{app:kernel}
%%==========================================================================

We derive Eq.~\eqref{eq:K-full} following the method of Papers~\cite{GhoshKlinkhamer2017} and \cite{Ghosh2026}.  Start from the vacuum-polarization kernel on the cylinder
$\R_\tau \times \R_x \times S^1_L$:
\begin{align}
  \pi_{\text{odd}}^{\mu\nu}&(p_r) \nonumber \\
  &= \frac{1}{L}\sum_{n=-\infty}^{\infty}\int\frac{\dd^2 l}{(2\pi)^2}\,
  \frac{\tr[\gamma^\mu(\slashed{l} + m_v)\gamma^\nu
    (\slashed{l} + \slashed{p} + m_v)]_{\text{odd}}}
       {(l_n^2 + m_v^2)((l_n + p_r)^2 + m_v^2)},
\label{eq:app-pi}
\end{align}
where $l_n = (\vec{l},\,(2\pi n + \vth_v)/L)$,
$p_r = (\vec{p},\,2\pi r/L)$, and $\slashed{l} = \gamma^\mu l_\mu$.
The subscript ``odd'' selects the parity-odd trace, which for
two-component fermions in $(2+1)$D is
$\tr[\gamma^\mu \gamma^\nu \gamma^\rho] = 2\epsilon^{\mu\nu\rho}$.

The mass-dependent odd contribution extracts the term linear in $m_v$
from the numerator.  After Feynman parameterization with parameter
$u \in [0,1]$ and shifting $l_\mu \to l_\mu - u p_\mu$, the
three-momentum integral gives
\begin{equation}
  \int\frac{\dd^2 l}{(2\pi)^2}\,
  \frac{1}{(l^2 + \Delta_{v,u}^2 + \omega_{n,u}^2)^2}
  = \frac{1}{4\pi(\Delta_{v,u}^2 + \omega_{n,u}^2)},
\label{eq:app-l-integral}
\end{equation}
where $\Delta_{v,u}^2 = m_v^2 + u(1-u)\tilde{p}^2$ and
$\omega_{n,u} = (2\pi n + \vth_v)/L + u(2\pi r/L)
= (2\pi n + \vth_v + 2\pi r u)/L$.  The parity-odd piece of the
kernel is therefore
\begin{align}
  \mathcal{K}_v^{\text{IR}}(p;L,\vth_v)
  &= \frac{\chi_v m_v}{2}\int_0^1 \dd u\,
  \frac{1}{L}\sum_{n=-\infty}^{\infty}
  \frac{1}{\Delta_{v,u}^2 + \omega_{n,u}^2}.
\label{eq:app-K-sum}
\end{align}
The Matsubara-type sum is evaluated using the standard identity
\begin{equation}
  \frac{1}{L}\sum_{n=-\infty}^{\infty}
  \frac{1}{\Delta^2 + \left(\frac{2\pi n + \theta}{L}\right)^2}
  = \frac{1}{2\Delta}\,
    \frac{\sinh(L\Delta)}{\cosh(L\Delta) - \cos\theta}.
\label{eq:app-Matsubara}
\end{equation}
%%%%%%%%%%%%%%
To prove Eq.~\eqref{eq:app-Matsubara}, define
$f(z) = [\Delta^2 + ((2\pi z + \theta)/L)^2]^{-1}$ and note that
$\frac{1}{L}\sum_{n=-\infty}^{\infty} f(n)$ equals the contour
integral $\oint_{\mathcal{C}} \dd z\,\pi\cot(\pi z)\,f(z)/(2\pi i L)$
over a contour enclosing the real axis.  Closing in the upper and
lower half-planes picks up the simple poles of $f(z)$ at
$z_{\pm} = (-\theta \pm iL\Delta)/(2\pi)$, giving
\begin{align}
  \frac{1}{L}\sum_{n=-\infty}^{\infty} f(n)
  &= \frac{1}{2\Delta}\left[
    \frac{1}{1 - e^{-L\Delta + i\theta}}
    + \frac{1}{1 - e^{-L\Delta - i\theta}}
  \right] \notag \\
  &= \frac{1}{2\Delta}\,
    \frac{2(e^{L\Delta} - \cos\theta)}
         {e^{2L\Delta} - 2e^{L\Delta}\cos\theta + 1} \notag \\
  &= \frac{1}{2\Delta}\,
    \frac{\sinh(L\Delta)}{\cosh(L\Delta) - \cos\theta},
  \label{eq:Matsubara-proof}
\end{align}
where the last step uses
$e^{2x} - 2e^x\cos\theta + 1 = 2e^x(\cosh x - \cos\theta)$.
Substituting with $\theta = \vth_v + 2\pi r u$ and $\Delta = \Delta_{v,u}$
immediately yields Eq.~\eqref{eq:K-full}.

%%==========================================================================
\section{Coleman-Hill non-renormalization}
\label{app:coleman-hill}
%%==========================================================================

We state the key steps of the Coleman-Hill~\cite{ColemanHill1985}
proof adapted to our setting.  Consider a $(2+1)$D gauge theory with
Lagrangian
\begin{align}
  \mathcal{L} = &-\frac{1}{2e_0^2}\partial^\mu A^\nu 
  (\partial_\mu A_\nu - \partial_\nu A_\mu) \nonumber \\ 
  &+ \frac{\mu}{2e_0^2}\epsilon^{\mu\nu\lambda}A_\mu\partial_\nu A_\lambda
  + \mathcal{L}_m(A_\mu, \phi^a),
\label{eq:CH-lagrangian}  
\end{align}
where $\mathcal{L}_m$ is the matter Lagrangian (scalars, spinors,
vectors) with gauge-invariant interactions.  The photon propagator
has the structure
\begin{align}
  iD_{\mu\nu}^{-1}(k)
  &= (k^2 g_{\mu\nu} - k_\mu k_\nu)\Pi_1(k^2) \nonumber \\
    &+ i\epsilon_{\mu\nu\lambda}k^\lambda \Pi_2(k^2)
    + \text{gauge terms},
\label{eq:CH-propagator}
\end{align}
and the topological mass is $\Pi_2(0)$.  At one loop,
\begin{equation}
  \Pi_2(0)\big|_{\text{1-loop}}
  = \frac{\mu}{e_0^2}
    + \frac{1}{4\pi}\sum_{\text{spinors}} q_a^2\,
      \frac{m_a}{|m_a|}.
\label{eq:CH-oneloop}
\end{equation}

\paragraph{Proof that higher loops vanish.}
The Ward identity $k_1^{\mu_1}\Gamma_{\mu_1\ldots\mu_n}^{(n)}
(k_1,\ldots) = 0$ together with analyticity of $\Gamma^{(n)}$ in all
momenta (guaranteed by massive matter) implies that for $n > 2$:
\begin{equation}
  \Gamma^{(n)}_{\mu_1\ldots\mu_n}(k_1,k_2,\ldots)
  = \mathcal{O}(k_1 k_2), \quad n > 2.
\label{eq:CH-ward}
\end{equation}
Consider a two-point self-energy graph at order $k$ in the external
momentum.  If the two external photon lines end on \emph{different}
matter loops, each factor $\Gamma^{(n)}$ with $n \geq 1$ carries at
least one power of $k$, giving $\mathcal{O}(k^2)$ overall and no
contribution to $\Pi_2(0)$.  If both end on the \emph{same} loop but
that loop has $n > 2$ photon lines attached, then
Eq.~\eqref{eq:CH-ward} gives $\mathcal{O}(k^2)$.  The only surviving
graph has exactly two external lines on a bare matter loop: this is
the one-loop diagram.

In our application: the BdG quasiparticles are the ``matter,'' the
probe field $a_\mu$ coupling to charge $2e$ or spin is the ``photon,''
and pair-pair interactions are included in $\mathcal{L}_m$.  All matter
fields are massive (gapped by $\Dpg$).  Therefore $\Pi_2(0)$,
which gives $\Kem$, is exact at one loop.  The gap $\Dpg(T)$ can be
renormalized by interactions, but the \emph{map from gap to CS level}
cannot.

%%==========================================================================
\section{Read-Green BdG Chern number}
\label{app:read-green}
%%==========================================================================

We review the derivation of the BdG Chern number following
Read and Green~\cite{Read2000}, Appendix~A.

For a spin-singlet or spin-triplet BdG system with unbroken U(1)
symmetry, the induced action for the external spin gauge field
$A_\mu$ is obtained by integrating out the Nambu fermions
$\Psi_{\bm{k}} = (c_{\bm{k}},\,i\sigma_y c_{-\bm{k}}^\dagger)^T/\sqrt{2}$.
The effective BdG Hamiltonian in Nambu space is
\begin{equation}
  K_{\text{eff}} = \sum_{\bm{k}}\Psi_{\bm{k}}^\dagger\,
  (\bm{E}_{\bm{k}} \cdot \bm{\sigma} \otimes I)\,\Psi_{\bm{k}},
\label{eq:RG-Keff}
\end{equation}
with $\bm{E}_{\bm{k}} = (\text{Re}\,\Delta_{\bm{k}},
-\text{Im}\,\Delta_{\bm{k}}, \xi_{\bm{k}})$ and quasiparticle
energy $E_{\bm{k}} = |\bm{E}_{\bm{k}}|$.  The Green's function is
\begin{equation}
  G^{-1}(p) = p_0 - \bm{E}_{\bm{p}} \cdot \bm{\sigma} \otimes I.
\label{eq:RG-Green}
\end{equation}

The spin response at linear order in the external field is determined
by the polarization tensor $\Pi_{\mu\nu}^{ab}(q)$.  Using the Ward
identity to dress both vertices at $q = 0$ (Read-Green,
Eq.~(\ref{eq:RG-Keff})), the quadratic induced action is the U(1)
Chern-Simons term
\begin{equation}
  S_{\text{ind}}[A] = \frac{\mathcal{M}}{4\pi}
  \int \dd^3 r\,\epsilon_{\mu\nu\lambda}\,
  A_\mu \partial_\nu A_\lambda,
\label{eq:RG-CS}
\end{equation}
where $\mathcal{M}$ is given by Eq.~\eqref{eq:RG-winding}.

For $p + ip$ pairing with $\Delta_{\bm{k}} =
\hat{\Delta}(k_x - ik_y)$ (the Read-Green $l=-1$
convention~\cite{Read2000}; the main text uses the conjugate
$l=+1$ convention $\Delta_{\bm{k}} \propto k_x + ik_y$, which
reverses the sign of $\mathcal{M}$ but not its magnitude):
as $|\bm{k}| \to 0$, $\bm{E}_{\bm{k}} \to (0,0,-\mu)$ if $\mu > 0$
(south pole) and $\to (0,0,+|\mu|)$ if $\mu < 0$ (north pole);
as $|\bm{k}| \to \infty$, $\bm{E} \to (0,0,+\infty)$ (north pole).
The winding number of the map $\hat{\bm{E}}: S^2 \to S^2$ is therefore
$+1$ for $\mu > 0$ (the map covers the full sphere) and $0$ for
$\mu < 0$ (the map misses the south pole).

This topological invariant depends only on $\sgn(\mu)$ and on the
angular momentum of $\Delta_{\bm{k}}$; it is independent of the
magnitude of $\Delta$ and of whether $\Delta = \Dsc$ or $\Dpg$.  The
BdG Chern number is therefore well-defined throughout the pseudogap
phase.

%%==========================================================================
\section{Gravitational Chern-Simons term and $\kappa_{xy}$}
\label{app:grav-cs}
%%==========================================================================

The relation between the gravitational CS coefficient and $\kxy$ is
derived by Ryu, Moore, and Ludwig~\cite{Ryu2012} and by
Volovik~\cite{Volovik2003}.  We state the essential steps.

A massive Dirac fermion coupled to a background metric (via the spin
connection $\omega_\mu^{ab}$) generates, upon integrating out the
fermion, a gravitational Chern-Simons term in the effective action.
For a single Dirac fermion of mass $m > 0$ in $(2+1)$D, the
one-loop calculation gives (Ryu-Moore-Ludwig, Sec.~III.C):
\begin{equation}
  I_{\text{CS}}^{\text{grav}}
  = \frac{1}{2}\cdot\frac{1}{4\pi}\cdot\frac{c}{24}
    \int \dd^3 x\,\epsilon^{ijk}\,\tr\!\left(
    \omega_i \partial_j \omega_k
    + \tfrac{2}{3}\omega_i \omega_j \omega_k\right),
\label{eq:grav-cs-detail}
\end{equation}
with $c = 1/2$ for a single Majorana fermion and $c = 1$ for a single
Dirac fermion.  For a BdG system with Chern number $C$ (counting the
number of chiral Majorana edge modes with $c_- = C/2$ per Majorana
species), the coefficient is $c_- = C \cdot (1/2)$ per spin species,
totaling $c_- = C$ for the physical system.

Luttinger's gravitational potential
formalism~\cite{Luttinger1964} identifies the thermal gradient
$\nabla T/T$ with a gravitational ``electric field''
$E_g = -c^2 \nabla T/T$, and the thermal Hall conductivity is
determined by the gravitational CS coefficient:
\begin{equation}
  \frac{\sigma_{xy}^T}{T} = c_-\,\frac{(\pi k_B)^2}{3h}
  = C\,\frac{\pi^2 k_B^2}{3h}.
\label{eq:sigma-T}
\end{equation}
Using the convention $\kxy \equiv \sigma_{xy}^T/T$ and
$\Kem = 2C$, this gives
\begin{equation}
  \kxy = \frac{\pi^2 k_B^2}{12h}\,\Kem
       = \frac{\pi^2 k_B^2}{6h}\,C.
\label{eq:kxy-from-grav}
\end{equation}
At finite temperature with gap $\Delta$, the replacement
$\Kem \to 2C\tanh(\Delta/2T)$ from Eq.~\eqref{eq:Kem-T} gives
Eq.~\eqref{eq:kxy-T}.

\end{document}